\documentstyle[epsfig]{l-aa}

\begin{document}

\def\ltsima{$\; \buildrel < \over \sim \;$}
\def\simlt{\lower.5ex\hbox{\ltsima}}

   \thesaurus{13.25.2 -- 11.19.1 -- 10.09.1}
   \title{BeppoSAX unveils the nuclear component in NGC 6240} 

   \author{P. Vignati 
          \inst{1}
   \and S. Molendi
          \inst{1}
   \and G. Matt
	  \inst{2}
   \and M. Guainazzi 
	  \inst{3}
   \and L.A. Antonelli
	  \inst{4}
   \and L. Bassani
          \inst{5}
   \and W.N. Brandt
          \inst{6}
   \and A.C. Fabian
          \inst{7}
   \and K. Iwasawa
          \inst{7}
   \and R. Maiolino
          \inst{8}
   \and G. Malaguti
          \inst{5}
   \and A. Marconi
          \inst{8}
   \and G.C. Perola
          \inst{2}
          }


   \institute{Istituto di Fisica Cosmica ``G.Occhialini'', Via Bassini 15,
        I--20133 Milano, Italy
   \and       Dipartimento di Fisica ``E. Amaldi",
              Universit\`a degli Studi Roma Tre,
              Via della Vasca Navale 84, I--00146 Roma, Italy
   \and Astrophysics Division, Space Science Department of
	ESA, ESTEC, Postbus 299, NL-2200 AG Noordwijk, The Netherlands 
      \and Osservatorio Astronomico di Roma, Via dell'Osservatorio,
        I--00044 Monteporzio-Catone, Italy
   \and Istituto Tecnologie e Studio Radiazioni Extraterrestri, CNR, Via
              Gobetti 101, I--40129 Bologna, Italy
   \and Dept. of Astronomy and Astrophysics, The Pennsylvania State
	University, 525 Davey Lab, 
	University Park, PA 16802, U.S.A.
     \and Institute of Astronomy,
        University of Cambridge, Madingley Road, Cambridge CB3 0HA, U.K.
    \and Osservatorio Astrofisico
             di Arcetri, L.~E.Fermi 5, I--50125 Firenze, Italy
}

   \date{To appear in A\&A Letters}

   \maketitle

\markboth{P. Vignati et al.}{BeppoSAX observation of NGC 6240}

   \begin{abstract}
The direct nuclear X--ray emission of the ultraluminous IRAS galaxy 
NGC~6240 has been observed for the first time by {\it Beppo}SAX. 
It is seen through an absorber with 
$N_{\rm H} \sim 2\times 10^{24}$ cm$^{-2}$.
The 2-10 keV X--ray luminosity of $>10^{44}$ erg s$^{-1}$ 
definitely proves that a powerful AGN is hosted
in this galaxy, despite the LINER optical appearance. 
Assuming the typical $L_{\rm X}/L_{\rm bol}$ ratio for quasars, 
our result implies that it is the AGN and not the starburst that 
dominates the energy output of NGC~6240. 
   \end{abstract}

%

\section{Introduction}

One of the major open problems with Ultraluminous Infrared Galaxies (ULIRG) 
is whether they host active nuclei at their centers. In at least a fraction
of objects, there is circumstantial evidence that this is the case. One
of the best examples in this respect is NGC~6240\footnote{If 
$L_{\rm IR} \geq 10^{12} L_{\odot}$ is the limiting criterion for ULIRG, 
strictly speaking NGC~6240 belongs to this class only if 
$H_0 \leq$65 km s$^{-1}$ Mpc$^{-1}$;
Genzel et al. (1998).}, a well known source  in which 
the presence of two nuclei (Fried \& Schulz 1983)
suggests an ongoing galaxy merger. Indications of the presence of an
AGN come mainly from X--rays: the flat 2--10 keV
X--ray spectrum and the prominent iron line complex observed by ASCA
(Iwasawa \& Comastri 1998) are
very similar to those of NGC~1068 (Iwasawa, Fabian \& Matt 1997; Matt et al. 
1997
and references therein) and suggest reflection from cold and warm
matter of an otherwise invisible nuclear component. On the other hand, 
in the optical spectrum all 
the diagnostic line ratios point towards a classification as a LINER 
(e.g. Veilleux et al. 1995), and the ISO SWS diagnostic diagram 
places NGC 6240 in the region of star formation dominance (Genzel et al.
1998). In soft X--rays the
spectrum is fairly complex, and best modeled by multi-temperature
plasma emission, very likely associated, at least partly, 
with the powerful starburst and related wind
present in this galaxy, a hypothesis strengthened by the ROSAT HRI
discovery of rather luminous ($\sim$10$^{42}$ erg cm$^{-2}$ s$^{-1}$)
emission extended over tens of kpc (Komossa, Schulz \& Greiner 1998; Schulz
et al. 1998). 
On the other hand, the point--like soft X--ray component 
is possibly due to scattering of the nuclear radiation (Komossa et al. 1998). 

A direct measurement of the nuclear luminosity is needed to assess the relative
importance of AGN and starburst emission. The only way to do that is to 
explore the hard X--ray band, which may reveal the direct 
nuclear component. If the absorbing matter is moderately thick to Compton
scattering (i.e. $N_{\rm H}\sim$a few$\times10^{24}$ cm$^{-2}$), the 
photons emerge above 10 keV, where the 
most sensitive instrument presently available is the PDS onboard 
{\it Beppo}SAX. This has 
already added two sources (Circinus Galaxy, Matt et al. 1999; Mrk~3,
Cappi et al. 1999) to the small list of ``moderately--thick" Seyfert 2s,
which was previously composed of only one object, NGC~4945 (Iwasawa et al.
1993; Done et al. 1996). 

\section{Observation and data reduction}

{\it Beppo}SAX  (Boella et al.  1997)  observed  NGC~6240 on 1998
August 14--17.
In this paper, data from  three  instruments are discussed: the LECS  
(0.1--10 keV),
the MECS (1.5--10 keV),
and the PDS (13--300 keV).

Data reduction and analysis follow the standard criteria, as described 
in Matt et al. (1997). 
The total  effective  exposure time was 52.5 ks in the LECS, 119.4 ks in the 
MECS and 114.4 ks in the PDS.
We  accumulated  and  examined  images for the MECS instruments.  
Comparing the radial profiles in three energy ranges (1.5--4,  4--7.5, 7.5--10
keV) with the corresponding Point Spread  Function, 
no evidence for extended emission is found. Light curves and spectra
were extracted within 
regions of 4$^{\prime}$  and of 2$^{\prime}$  radius  centered on the source 
for
MECS and LECS, respectively (the small LECS extraction radius is chosen to
avoid contamination from a nearby, soft source, not visible in
the MECS).  The light curves are  consistent  with  being  constant.  The
resulting  count rate is (8.46$\pm$0.67)$\times$10$^{-3}$ counts s$^{-1}$
(0.1--4.5 keV) for the LECS and (2.46$\pm$0.07)$\times$10$^{-2}$ 
counts s$^{-1}$ (1.5--10 keV) in the MECS.  
No significant variability was found in the PDS 
light--curve either, with a count rate of 0.38$\pm$0.04 counts s$^{-1}$.
We restricted the spectral analysis to the 0.1--4.5 keV and
1.5--10 keV energy bands for LECS and MECS respectively, where the latest
released  (September  1997)  response  matrices are best  calibrated.  Standard
blank--sky  files provided by the {\it  Beppo}SAX  Science Data Center (SDC) 
were used for the  background  subtraction. 

\section{Spectral analysis and results}
In Fig. 1 we compare  the $>$2 keV spectra of
NGC~6240  and NGC~1068.  Both spectra have been
normalized to the Crab spectrum; since the latter is a power law  with  
photon  index $\simeq$ 2, we are 
basically  showing the $\nu  F_{\nu}$  spectrum.
The MECS spectra of the two sources are similar in shape, not
surprisingly as we know that in this
energy range both sources have been successfully fitted 
by two reflection components,
warm and cold (e.g. Iwasawa et al. 1997; Iwasawa \& Comastri 1998).  
While below 10 keV the flux of NGC~6240 is the lowest, the opposite occurs
in the PDS band. Clearly, in NGC~6240 a new component emerges at high
energies. No known bright sources are present in the field of view of the 
PDS, and the probability of a serendipitous source in 
the $1.3^{\circ}$ PDS field with a flux equal or larger than
that of NGC~6240 is $\simlt 10^{-5}$, if the 2--10
keV ASCA  LogN--LogS  (Cagnoni et al. 1998) is adopted, and a power law
spectrum with photon index 1.8 is assumed. 
Thus the best explanation for this component is 
nuclear emission piercing through an absorber with column 
density $N_H$ a few$\times$10$^{24}$ cm $^{-2}$, i.e. Compton--thick 
but still permitting partial transmission above $\sim$10 keV. 
In Fig.~2, the broad band spectrum is fitted
following Iwasawa \& Comastri (1998): an excess is evident in the PDS band.
We therefore modeled the continuum  above 2 keV with:
{\it i}) a power law, representing the intrinsic emission, with a 
photon index fixed 
at a typical value of  $\Gamma$=1.8, absorbed
by a column  density allowed  to be free (leaving the index 
as a free parameter, a value consistent with the adopted one,
but very loosely constrained,  is found); {\it ii}) a second, unabsorbed
power law with the same index, 
standing for the warm  reflection  component; {\it iii}) a pure cold  
reflection
component (PEXRAV model in XSPEC) generated by a power law with the same 
index.

\begin{figure}[htb]
\centerline{\hspace{-0.7cm}
\epsfig{figure=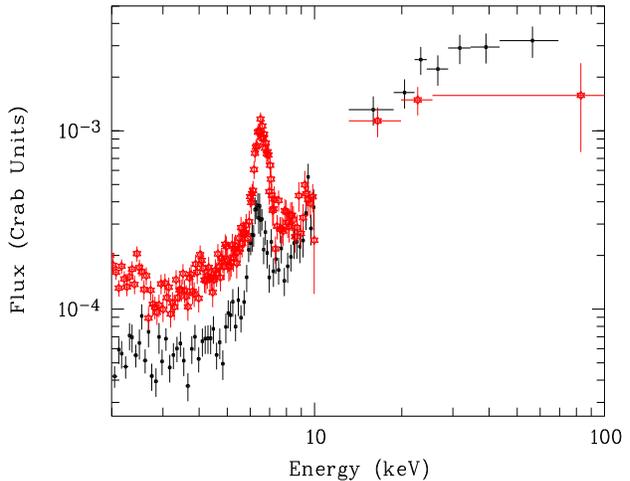,height=6.8cm, width=9.0cm, angle=270}} 
\vspace{-0.5cm}
\caption[]{Comparison of the MECS and PDS spectra of NGC 1068 (open 
stars) with those of NGC 6240 (dots). Both spectra 
have been normalized 
to the Crab spectrum, which is a power--law with a photon index of $\sim$ 2.}
\end{figure}

As we are interested here in the high energies, and because below 
2 keV ASCA is superior to {\it Beppo}SAX as far as both sensitivity and
spectral resolution are concerned, we content ourselves with adopting
 in this band
the best fitting model of Iwasawa \&  Comastri (1998). It consists of  
a  multi-temperature  optically  thin plasma, which  probably  originates in a
powerful  starburst.  In particular we use a two--temperature  MEKAL \
model, where
the absorption of the coldest MEKAL was fixed at the Galactic value 
(N$_{\rm H}$ = 5.8$\times$  10$^{20}$  cm$^{-2}$),  while the  absorption  
of the other  MEKAL was
allowed to be a free parameter.  We applied the latter  absorption  to
all the AGN  components.  The  envisaged  scenario  is then 
one in which  an  external
starburst  region,  associated  with the cooler  MEKAL  component,  absorbs  
emission from the inner regions of NGC~6240 (e.g. Fabian el al.  1998).

A prominent and broad iron K line complex,  very 
similar to that seen in NGC~1068, is also evident.
Because of this  similarity we assume that 
the line is actually a blend of: the 6.4 keV K$\alpha$ and 7.06 keV K$\beta$  
fluorescent lines from cold iron (the intensity
of the latter has been forced to be equal to 1/9 that of the former, 
as expected from atomic physics); the 6.7 and 6.97 K$\alpha$  recombination 
lines from He-- and H--like iron. 
In summary the complete model, $F$, we used to fit our data can be written as:
$$ F = A_{\rm G}\{A_{\rm SB}[A_{\rm T}(PL) + R_{\rm W} + R_{\rm C} + M_{\rm H} 
+ 4GL] + M_{\rm C}\},$$
where: $A_{\rm G}$ is the absorption associated with the Galactic column,
$A_{\rm SB}$ is the absorption related to the starburst, 
$A_{\rm T}$ is the absorption acting on the nuclear emission (including
the Compton cross section), possibly 
associated with the molecular torus shrouding the nucleus,
$PL$ is the power--law modeling the nuclear component,
$R_{\rm W}$ and $R_{\rm C}$ are respectively the warm (optically thin)
and cold (optically thick) reflection components,
$M_{\rm H}$ and $M_{\rm C}$ are respectively the cold and hot thermal components
and $4GL$ are the four Gaussian lines modeling the iron blend.

\vspace{-0.15in}
\begin{table}
\caption{Best fit parameters, see text for details. }
\begin{tabular}{|c|c|c|}
\hline
Component & Parameter & Value \cr
\hline
~ & ~ & ~\cr
$A_{\rm SB}$ & N$_{\rm H}$~(10$^{22}$~cm$^{-2}$) & 1.99$^{+1.23}_{-0.67}$ \cr
$A_{\rm T}$ & N$_{\rm H}$~(10$^{22}$~cm$^{-2}$) & 218$^{+40}_{-27}$ \cr
$PL$ & $\Gamma$ & 1.8~(fixed) \cr
~ & Flux~at~1~ keV$^a$ & 4.1($^{+1.3}_{-1.3}$)$\times10^{-2}$ \cr
$GL1$ & E$_{line}$ & 6.4~(fixed)   \cr
~ & I$_{line}$$^b$ & 2.23($^{+0.56}_{-0.51}$)$\times10^{-5}$ \cr
$GL2$ & E$_{line}$ & 6.7~(fixed)  \cr
~ & I$_{line}$$^b$ & 0.70($^{+0.74}_{-0.69}$)$\times10^{-5}$ \cr
$GL3$ & E$_{line}$ & 6.97~(fixed)   \cr
~ & I$_{line}$$^b$ & 0.53($^{+0.45}_{-0.43}$)$\times10^{-5}$ \cr
$GL4$ & E$_{line}$ & 7.06~(fixed)  \cr
~ & I$_{line}$$^b$ & 0.11$\times$I$_{line}$(6.4~keV) \cr
$R_{\rm W}$ & $\Gamma$ & 1.8~(fixed)  \cr
~ & Flux~at~1~ keV$^a$ & 0.24($^{+0.09}_{-0.09}$)$\times10^{-3}$ \cr
$R_{\rm C}$ & $\Gamma$ & 1.8~(fixed)  \cr
~ & Flux~at~1~ keV$^a$ & 2.4($^{+1.1}_{-1.2}$)$\times10^{-3}$ \cr
$M_{\rm H}$ & $\rm kT$~(keV) & 0.51$^{+0.14}_{-0.18}$ \cr
~ & Flux$^c$ & 9.4($^{+4.1}_{-2.2}$)$\times10^{-13}$ \cr
$M_{\rm C}$ & $\rm kT$~(keV) & 0.70$^{+0.19}_{-0.25}$ \cr
~ & Flux$^c$ & 1.3($^{+5.1}_{-0.7}$)$\times10^{-11}$ \cr
~ & ~ & ~\cr 
\hline
\end{tabular}
~\par
$^a$Photons cm$^{-2}$ s$^{-1}$ keV$^{-1}$; 
$^b$Photons cm$^{-2}$ s$^{-1}$; 
$^c$Bolometric flux in erg cm$^{-2}$ s$^{-1}$
\end{table}

In Fig. 3 we show the deconvolved best fitting model with all the 
components. The best fit parameters are listed in Table 1.
The fit is acceptable, ${\chi}^2$/d.o.f.=131/119.  
If we do not include the transmission component, 
and try therefore to account for the spectral
component emerging in the PDS band with pure Compton reflection (as
in NGC1068, Matt et al. 1997), the fit is significantly worse 
(${\rm \chi^2 =179/121}$~dof; see Fig.~2). 
It must be noted that the 
absorber of the nuclear emission, $A_{\rm T}$, is Compton--thick. A
proper model, including Compton scattering within the matter 
(see Matt et al. 1999 and Matt, Pompilio \& La Franca 1999 for details
on this model)
should have then been used, if this absorber has a significant 
covering factor (as for example for a torus geometry with no significant
vertical density gradient). When this is done,
however, the parameters of the various components 
do not differ significantly from those obtained with the simpler model, apart
from the flux associated with the intrinsic emission, which turns out to be 
about one third that reported in Table 1. This is because a fraction
of the photons
emitted towards other directions are now scattered into the line of sight, 
and the nuclear luminosity required to explain the observed flux is 
therefore lower. 

The same fit but without $R_{\rm C}$ is also acceptable
(${\rm \chi^2 = 133/120}$~dof). In this case, the observed 6.4 keV iron line
would entirely originate in the thick absorbing medium, which is possible
provided that the covering factor is not much smaller than unity, as we
have verified by Monte Carlo simulations. Therefore, the presence
of the cold reflection component is not certain in this source. 
If $R_{\rm W}$ is instead excluded, the
fit is significantly worse (${\rm \chi^2 = 139/120}$~dof).

\begin{figure}[htb]       
\centerline{\epsfig{figure=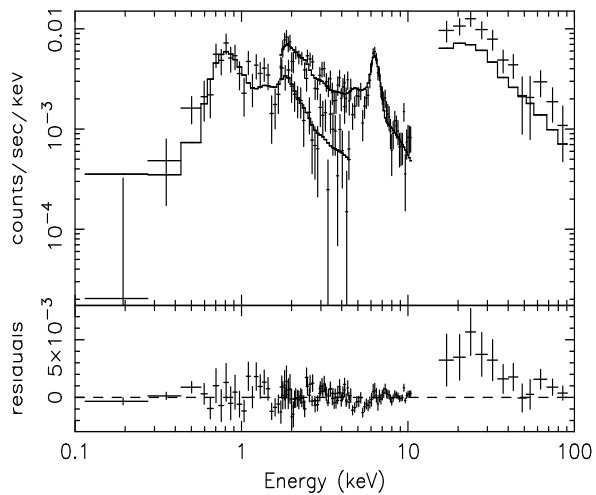, height=7.0cm,  width=8.3cm,   angle=270}}
\vspace{-0.5cm}
\caption[]{LECS, MECS and PDS spectra of NGC 6240 and best fitting model
and residuals, when the Iwasawa \& Comastri (1998) model is adopted. 
A further component is clearly emerging in the PDS band. } 
\end{figure}

\begin{figure}[htb]       
\centerline{\epsfig{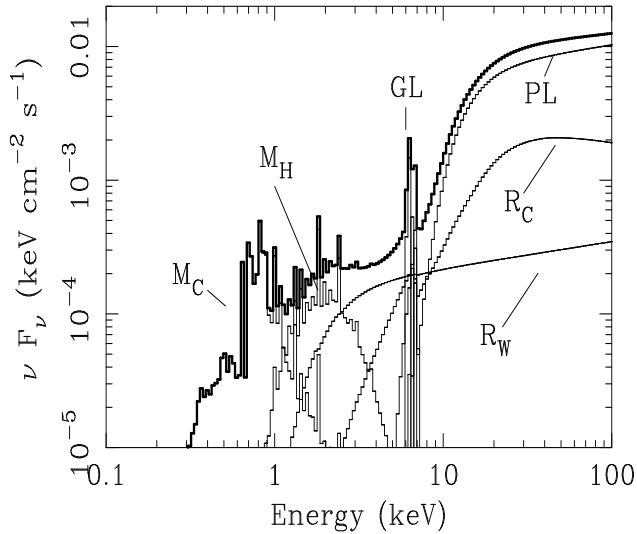}}  
\caption[]{Deconvolved best fitting model for NGC 6240. The thick line
represents the total model, the thin lines the spectral components,
labeled as in Table 1.
 } 
\end{figure}

\section{Discussion}

The main result of the {\it Beppo}SAX observation of NGC~6240
is the discovery and measurement of the nuclear emission. 
The 2--10 keV nuclear  luminosity,  after  correction  for  absorption, 
is 3.6$\times$10$^{44}$ erg s$^{-1}$ (assuming $H_0$=50 km s$^{-1}$ 
Mpc$^{-1}$),
or about one
third of this if the absorber covers a large solid angle, like in a torus. 

As discussed above, the presence of cold reflection is not unambiguously
established in this source. Let us, however,
assume that it does exist, and take the best 
fitting parameters listed in Table~1. The ratio between the 2--10 keV 
cold  reflection  and  intrinsic fluxes is 
about 0.3\% for the higher value of the intrinsic luminosity and 3 
times larger for the lower value.  
Assuming a torus  geometry  like that  adopted by  Ghisellini, Haardt \& Matt 
(1994), and consistently the lower figure for the intrinsic luminosity,
we derive an inclination  angle  of the  torus  
axis  with  respect  to the  observer  of
$\sim$35$^{\circ}$ ($\sim$45$^{\circ}$ if the upper figure is instead 
adopted).
It is worth noting that both NGC 6240 and Circinus Galaxy (Guainazzi et al. 
1999) appear
to be observed at moderate inclinations, while NGC 1068, which is
totally obscured, is possibly seen almost edge-on (Greenhill et al. 1996, 
Matt et al. 1997; but see Kishimoto 1999 for a different point of view). 
This would suggest a decrease in
the density of the X--ray absorbing medium with 
increasing distance from the equatorial plane.

The ratio between 2--10 keV warm reflection  and  intrinsic 
fluxes is about 0.5\% (1.5\% for the lower luminosity). 
The He-- and H--like lines
have EW with respect to the warm reflection component of 
about 1 and 0.8 keV, respectively. From  these values we 
estimate the column  density of the warm mirror (e.g. Matt, Fabian \&
Brandt 1996; Guainazzi et al. 1999), 
$N_{\rm H,warm}$, to be between 10$^{21}$ and  10$^{22}$  atoms
cm$^{-2}$, roughly consistent with the value estimated from the ratio 
between the normalizations of the warm reflection and of
the intrinsic emission. 




The 2-10 keV X--ray luminosity is 10 to 30 times 
lower than the IR luminosity (calculated as prescribed by Mulchaey et al. 
1994), consistent with the usual values for 
Seyferts  (e.g. Mulchaey et al. 1994; Awaki 1997).
In the case of NGC~6240, therefore, a ULIRG
turns out by no means to be a peculiar AGN, as far as X--ray 
properties are concerned. 
If $L_{\rm 2-10~keV}/L_{\rm bol}\sim$0.03, as typical for QSO (Elvis et al. 1994),
 our result implies that it is the AGN that dominates the 
energy output, and not the starburst as deduced by 
Genzel et al. (1998) on the basis of the ISO spectrum. 

In the optical, NGC~6240 is classified as a LINER. Possibly, 
the NLR is obscured by the starburst region, whose column 
density, from our analysis, corresponds to A$_{\rm V}\sim$10. 
It is worth noting that NGC 6240 is not the first LINER that,
when observed in X-rays, turns out to be an AGN (see e.g., NGC 4945, 
Done et al. 1996; NGC 1052, Guainazzi \& Antonelli 1999). Differently
from them (but similarly to two other ULIRG, IRAS~23060+0505: Brandt et al. 
1997 and IRAS~20460+1925: Ogasaka et al. 1997), NGC~6240 has a high X--ray 
luminosity, and it may be called
an obscured QSO. While it would be premature to draw conclusions
on ULIRGs as a class, one cannot help noting that 
the so far elusive type 2 QSO should be better searched for 
in X--rays rather than in the optical band. 

\begin{acknowledgements}
We thank the referee, R. Antonucci, for valuable comments.
We acknowledge the BeppoSAX SDC team for providing pre--processed event files
and the support in data reduction.  GM and GCP acknowledge financial support 
from ASI, WNB from NASA LTSA grant NAG5-8107. 
MG acknowledges an ESA Research Fellowship. BeppoSAX is a joint Italian-Dutch
program.

\end{acknowledgements}


\begin{thebibliography}{} 


\bibitem[]{}  Awaki H., 1997, in ``Emission Lines in Active Galaxies: New 
Methods and Techniques", ASP conf. series Vol 113, p. 44

\bibitem[]{} Boella G., Butler R.C., Perola G.C., et al., 1997, A\&AS 112, 299



\bibitem[]{} Brandt W.N., Fabian A.C., Takahashi K., Fujimoto R., Yamashita
A., Inoue H., Ogasaka Y., 1997, MNRAS 290, 617

\bibitem[]{} Cagnoni I., Della Ceca R., Maccacaro T., 1998, ApJ 493, 54

\bibitem[]{} Cappi M., Bassani L., Comastri A.,  et al., 1999, A\&A 344, 857

\bibitem[]{} Done C., Madejski G.M., Smith D.A., 1996, ApJ 463, 63

\bibitem{} Fabian A.C., Barcons X., Almaini O., Iwasawa K., 1999, MNRAS 297, 
L11

\bibitem{} Fried J.W., Schulz H., 1983, A\&A 118, 166


\bibitem[]{} Genzel  R., Lutz D., Sturm E., et al., 1998,  A\&A 498, 579

\bibitem[]{} Ghisellini G., Haardt F., Matt G., 1994, MNRAS 267, 743

\bibitem[]{} Greenhill, L. J., Gwinn, C. R., Antonucci, R., Barvainis, R.,
1996, ApJ 472, L21

\bibitem[]{} Guainazzi M., Matt G., Antonelli L.A., et al., 1999, MNRAS in 
press
(astro-ph/9905261)

\bibitem[]{} Guainazzi M., Antonelli, L.A., 1999, MNRAS 304, L15

\bibitem[]{} Iwasawa K., Koyama K., Awaki H., et al., 1993, ApJ 409, 155

\bibitem[]{} Iwasawa K., Fabian A.C., Matt G., 1997, MNRAS 289, 443

\bibitem[]{} Iwasawa K., Comastri A., 1998, MNRAS 297, 1219

\bibitem[]{} Kishimoto M., 1999, ApJ 518, 676

\bibitem[]{} Komossa S., Schulz H., Greiner J., 1998, A\&A 334, 110

\bibitem[]{} Matt G., Brandt W.N,, Fabian A.C., 1996, MNRAS 280, 823 

\bibitem[]{}  Matt G., Guainazzi M., Frontera F., et al., 1997, A\&A 325, L13

\bibitem[]{}  Matt G., Pompilio F., La Franca F., 1999, New Astronomy 4/3, 191

\bibitem[]{}  Matt G., Guainazzi M., Maiolino R., et al., 1999, A\&A 341, L39

\bibitem[]{}  Mulchaey J.S., Koratkar A., Ward M.J., 1994, ApJ 436, 586

309

\bibitem[]{} Ogasaka Y., Inoue H., Brandt W.N., et al., 1997, PASJ 49, 179

\bibitem[]{} Schulz H., Komossa S., Berg\"{o}fer T.W., Boer B., 1998, 
A\&A 330, 823

\bibitem[]{} Veilleux S., Kim D.-C., Sanders D.B., Mazzarella J.M., Soifer
B.T., 1995, ApJSS 98, 171

\end{thebibliography}
\end{document}